\documentstyle[twoside]{article}

\catcode`\@=11
\long\def\@makefntext#1{ 
\protect\noindent \hbox to 3.2pt {\hskip-.9pt
$^{{\eightrm\@thefnmark}}$\hfil}#1\hfill} 

\def\thefootnote{\fnsymbol{footnote}}
 \def\@makefnmark{\hbox to 0pt{$^{\@thefnmark}$\hss}}  

\def\ps@myheadings{\let\@mkboth\@gobbletwo
\def\@oddhead{\hbox{} 
\rightmark\hfil\eightrm\thepage}
\def\@oddfoot{}\def\@evenhead{\eightrm\thepage\hfil 
\leftmark\hbox{}}\def\@evenfoot{}
\def\sectionmark##1{}\def\subsectionmark##1{}}

\oddsidemargin=\evensidemargin
\renewcommand{\thefootnote}{\fnsymbol{footnote}}

\newcounter{sectionc}\newcounter{subsectionc}\newcounter{subsubsectionc}
\renewcommand{\section}[1] {\vspace{12pt}\addtocounter{sectionc}{1}
\setcounter{subsectionc}{0}\setcounter{subsubsectionc}{0}\noindent
	{\tenbf\thesectionc. #1}\par\vspace{5pt}}
\renewcommand{\subsection}[1] {\vspace{12pt}\addtocounter{subsectionc}{1}
	\setcounter{subsubsectionc}{0}\noindent
	{\bf\thesectionc.\thesubsectionc. {\kern1pt \bfit #1}}\par\vspace{5pt}}
\renewcommand{\subsubsection}[1] {\vspace{12pt}\addtocounter{subsubsectionc}{1}
	\noindent{\tenrm\thesectionc.\thesubsectionc.\thesubsubsectionc.
	{\kern1pt \tenit #1}}\par\vspace{5pt}}
\newcommand{\nonumsection}[1] {\vspace{12pt}\noindent{\tenbf #1}
	\par\vspace{5pt}}

\newcounter{appendixc}
\newcounter{subappendixc}[appendixc]
\newcounter{subsubappendixc}[subappendixc]
\renewcommand{\thesubappendixc}{\Alph{appendixc}.\arabic{subappendixc}}
\renewcommand{\thesubsubappendixc}
	{\Alph{appendixc}.\arabic{subappendixc}.\arabic{subsubappendixc}}

\renewcommand{\appendix}[1] {\vspace{12pt}
        \refstepcounter{appendixc}
        \setcounter{figure}{0}
        \setcounter{table}{0}
        \setcounter{lemma}{0}
        \setcounter{theorem}{0}
        \setcounter{corollary}{0}
        \setcounter{definition}{0}
        \setcounter{equation}{0}
        \renewcommand{\thefigure}{\Alph{appendixc}.\arabic{figure}}
        \renewcommand{\thetable}{\Alph{appendixc}.\arabic{table}}
        \renewcommand{\theappendixc}{\Alph{appendixc}}
        \renewcommand{\thelemma}{\Alph{appendixc}.\arabic{lemma}}
        \renewcommand{\thetheorem}{\Alph{appendixc}.\arabic{theorem}}
        \renewcommand{\thedefinition}{\Alph{appendixc}.\arabic{definition}}
        \renewcommand{\thecorollary}{\Alph{appendixc}.\arabic{corollary}}
        \renewcommand{\theequation}{\Alph{appendixc}.\arabic{equation}}
        \noindent{\tenbf Appendix \theappendixc #1}\par\vspace{5pt}}
\newcommand{\subappendix}[1] {\vspace{12pt}
        \refstepcounter{subappendixc}
        \noindent{\bf Appendix \thesubappendixc. {\kern1pt \bfit #1}}
	\par\vspace{5pt}}
\newcommand{\subsubappendix}[1] {\vspace{12pt}
        \refstepcounter{subsubappendixc}
        \noindent{\rm Appendix \thesubsubappendixc. {\kern1pt \tenit #1}}
	\par\vspace{5pt}}

\newcommand{\textlineskip}{\baselineskip=13pt}
\newcommand{\smalllineskip}{\baselineskip=10pt}

\def\eightcirc{
\begin{picture}(0,0)
\put(4.4,1.8){\circle{6.5}}
\end{picture}}
\def\eightcopyright{\eightcirc\kern2.7pt\hbox{\eightrm c}}

\def\abstracts#1#2#3{{
	\centering{\begin{minipage}{4.5in}\baselineskip=10pt\eightrm
	\centerline{ABSTRACT}
	\parindent=0pt #1\par
	\parindent=15pt #2\par
	\parindent=15pt #3
	\end{minipage} }\par}}

\newcommand{\bibit}{\nineit}

\renewenvironment{thebibliography}[1]			
	{\ninerm
	 \baselineskip=11pt				
	 \begin{list}{\arabic{enumi}.}
	{\usecounter{enumi}\setlength{\parsep}{0pt}
	 \setlength{\leftmargin 17pt}{\rightmargin 0pt}	
	 \setlength{\itemsep}{0pt} \settowidth		
	{\labelwidth}{#1.}\sloppy}}{\end{list}}

\newcounter{itemlistc}
\newcounter{romanlistc}
\newcounter{alphlistc}
\newcounter{arabiclistc}

\newenvironment{romanlist}
	{\setcounter{romanlistc}{0}
	 \begin{list}{$($\roman{romanlistc}$)$}
	{\usecounter{romanlistc}
	 \setlength{\parsep}{0pt}
	 \setlength{\itemsep}{0pt}}}{\end{list}}

\newcommand{\fcaption}[1]{
        \refstepcounter{figure}
        \setbox\@tempboxa = \hbox{\eightrm Fig.~\thefigure. #1}
        \ifdim \wd\@tempboxa > 5in
           {\begin{center}
        \parbox{5in}{\eightrm \smalllineskip Fig.~\thefigure. #1 }
            \end{center}}
        \else
             {\begin{center}
             {\eightrm Fig.~\thefigure. #1}
              \end{center}}
        \fi}

\newcommand{\tcaption}[1]{
        \refstepcounter{table}
        \setbox\@tempboxa = \hbox{\eightrm Table~\thetable. #1}
        \ifdim \wd\@tempboxa > 5in
           {\begin{center}
        \parbox{5in}{\eightrm\smalllineskip Table~\thetable. #1 }
            \end{center}}
        \else
             {\begin{center}
             {\eightrm Table~\thetable. #1}
              \end{center}}
        \fi}

\def\@citex[#1]#2{\if@filesw\immediate\write\@auxout	
	{\string\citation{#2}}\fi			
\def\@citea{}\@cite{\@for\@citeb:=#2\do			
	{\@citea\def\@citea{,}\@ifundefined		
	{b@\@citeb}{{\bf ?}\@warning
	{Citation `\@citeb' on page \thepage \space undefined}}
	{\csname b@\@citeb\endcsname}}}{#1}}

\newif\if@cghi
\def\cite{\@cghitrue\@ifnextchar [{\@tempswatrue
	\@citex}{\@tempswafalse\@citex[]}}
\def\citelow{\@cghifalse\@ifnextchar [{\@tempswatrue
	\@citex}{\@tempswafalse\@citex[]}}
\def\@cite#1#2{{$\null^{#1}$\if@tempswa\typeout
	{IJCGA warning: optional citation argument
	ignored: `#2'} \fi}}

\def\pmb#1{\setbox0=\hbox{#1}
	\kern-.025em\copy0\kern-\wd0
	\kern.05em\copy0\kern-\wd0
	\kern-.025em\raise.0433em\box0}

\def\fnt#1#2{\footnotetext{\kern-.3em
	{$^{\mbox{\scriptsize #1}}$}{#2}}}

\def\fpage#1{\begingroup
\voffset=.3in
\thispagestyle{empty}\begin{table}[b]\centerline{\footnotesize #1}
	\end{table}\endgroup}

\def\runninghead#1#2{\pagestyle{myheadings}
\markboth{{\eightit{\quad #1}}\hfill}{\hfill{\eightit{#2\quad}}}}

\font\tenbf=cmbx10
\font\tenit=cmti10
\font\tenit=cmti10
 1
 1
 1

\font\ninerm=cmr9
\font\nineit=cmti9

\font\eightrm=cmr8
\font\eightit=cmti8

\newcommand{\overscript}[2]%
    {%
     \begin{array}[b]{@{}c@{}} {\scriptscriptstyle #2} \\ #1 \end{array}}

\renewcommand{\overscript}[2]%
    {%
     \begin{array}[b]{@{}c@{}}%
     {\raisebox{0pt}[0pt][0pt]{$\scriptscriptstyle #2$}} \\ #1 \end{array}}

\newcommand{\underscript}[2]%
    {%
     \begin{array}[t]{@{}c@{}} #1 \\ {\scriptscriptstyle #2}  \end{array}}

\def\qed{\hbox{${\vcenter{\vbox{                          
   \hrule height 0.4pt\hbox{\vrule width 0.4pt height 6pt
   \kern5pt\vrule width 0.4pt}\hrule height 0.4pt}}}$}}

\runninghead{Consistent interactions and local BRST cohomology
$\ldots$} {Consistent interactions and local BRST cohomology
$\ldots$}

\begin{document}
\normalsize\textlineskip
{\thispagestyle{empty}
\setcounter{page}{1}

\renewcommand{\thefootnote}{\fnsymbol{footnote}} 

\hspace{9cm} ULB-PMIF/93-04
\fpage{1}
\vskip 2 cm
\centerline{\bf CONSISTENT INTERACTIONS BETWEEN GAUGE FIELDS}
\vspace*{0.035truein}
\centerline{\bf AND LOCAL BRST COHOMOLOGY :}
\vspace*{0.035truein}
\centerline{\bf THE EXAMPLE OF YANG-MILLS MODELS\footnote{Talk given by
G. Barnich at the
``Journ\'ees Relativistes, Bruxelles '93".}}
\vspace{0.37truein}
\centerline{\footnotesize G. BARNICH\footnote{Aspirant au Fonds
National de la Recherche Scientifique (Belgium).}, M. HENNEAUX\footnote{Also
at Centro de Estudios Cient\'{\i}ficos de Santiago, Casilla 16443,
Santiago 9, Chile.} and R. TATAR\footnote{Permanent address :
Faculty of Science,
University of Craiova, Alexandru Ioan Cuza street 13, 1100 Craiova, Romania.}}
\vspace*{0.015truein}
\centerline{\footnotesize\it Facult\'e des Sciences, Universit\'e Libre de
Bruxelles,}
\baselineskip=10pt
\centerline{\footnotesize\it Campus Plaine C.P. 231, B-1050 Bruxelles, Belgium}

\vspace*{0.21truein}

\abstracts{\noindent Recent results on the cohomological reformulation of the
problem
of consistent interactions between gauge fields are illustrated in the case of
the Yang-Mills models.
By evaluating the local BRST cohomology through descent equation techniques,
it is shown (i) that there is a unique local, Poincar\'e invariant
cubic vertex for free gauge vector fields which preserves the number of
gauge symmetries to first order in the coupling constant; and (ii) that
consistency to second order
in the coupling constant requires the structure constants appearing in the
cubic vertex to
fulfill the Jacobi identity.
The known uniqueness of the Yang-Mills coupling is therefore rederived through
cohomological arguments.}{}{}
\vspace*{-3pt}\textlineskip
\textheight=7.8truein
\setcounter{footnote}{0}
\renewcommand{\thefootnote}{\alph{footnote}}

\section{Introduction}
\noindent
Consider a free gauge theory with action
${\overscript{S}{(0)}}_0[\varphi^i]$
and gauge symmetry given by
\begin{equation}
\delta_\varepsilon \varphi^i = \overscript{R^i_\alpha}{(0)}
\varepsilon^\alpha,\ \
{\delta {\overscript{S}{(0)}}_0 \over \delta\varphi^i}
{\overscript{R^i_\alpha}{(0)}}=0.
\end{equation}
The gauge symmetry removes unphysical degrees of freedom which would otherwise
either make the
theory unstable or introduce negative norm states. The question investigated
here is whether one
can introduce couplings among the fields $\varphi^i$ which fulfill the crucial
physical
requirement of preserving the number of gauge
symmetries\cite{Arnowitt,Berends,Bengtsson}.
Interactions fulfilling this
condition will be said to be ``consistent" [other consistency requirements
such as causal physical
propagation may have to be imposed, but this question will not be studied
here].

In a recent paper\cite{Barnich} it has been shown that the problem of
constructing consistent
interactions can be naturally reformulated
as a deformation problem\cite{Gerstenhaber} of the solution of the master
equation. This is so because
the master equation contains all the information about the
gauge structure of the theory\cite{Batalin,Henneaux}. If a consistent
interacting gauge theory
can be constructed, then the solution $\overscript{S}{(0)} =
{\overscript{S}{(0)}}_0 +
``antifield\  contributions"$ of the master equation for the free
theory can be deformed into a solution $S$,
\begin{eqnarray}
&S = \overscript{S}{(0)} +
g \overscript{S}{(1)} +
g^2 \overscript{S}{(2)} + ...\\
&(S,S)=0\label{mastereq},
\end{eqnarray}
of the master equation for the deformed theory which has the same spectrum of
ghosts and antifields.

Eq.(\ref{mastereq}) can be
analyzed order by order in the deformation parameter $g$
(= coupling constant), leading to
\begin{eqnarray}
(\overscript{S}{(0)},\overscript{S}{(0)})&= 0 \label{deformation1}\\
2(\overscript{S}{(0)},\overscript{S}{(1)})&= 0 \label{deformation2}\\
2(\overscript{S}{(0)},\overscript{S}{(2)}) +
(\overscript{S}{(1)},\overscript{S}{(1)})&= 0 \label{deformation3}\\
&\vdots\ \ \ .\nonumber
\end{eqnarray}
Eq. (\ref{deformation1}) is the master equation for the given
free gauge theory and is thus satisfied. Eq. (\ref{deformation2}) requires
$\overscript{S}{(1)}$
to be a cocycle of the free BRST differential $\overscript{s}{(0)}\equiv
(\cdot,\overscript{S}{(0)})$. Only cohomologically non trivial solutions of
(\ref{deformation2})
have to be considered, because BRST exact solutions $\overscript{S}{(1)}=
g(K,\overscript{S}{(1)})$ correspond to interactions obtained from the free
theory by making non-linear field redefinitions. Hence $\overscript{S}{(1)}$
belongs to
$H^0(\overscript{s}{(0)})$, the ghost number zero cohomological space of
$\overscript{s}{(0)}$.
This space is known\cite{Henneaux} to be isomorphic with the space of physical
observables
of the free theory, i.e., the equivalence classes of on- shell gauge invariant
functionals
that coincides when the equations of motion of the free theory hold.

Turn now to eq.(\ref{deformation3}).
Under very general regularity assumptions, the antibracket can be
shown\cite{Barnich,Barnich1}
to be to be trivial in cohomology, i.e., the antibracket of two BRST closed
functionals is
BRST exact. This means that $(\overscript{S}{(1)},\overscript{S}{(1)})$
is BRST exact, so that eq.(\ref{deformation3}) represents no obstruction
to continuing the construction of the interacting action and that
$\overscript{S}{(2)}$ exists. Similarily, one finds that the higher order
equations
can also be satisfied
and that the construction of an interacting theory starting from an element
$\overscript{S}{(1)}\ \in H^0(\overscript{s}{(0)})$ is unobstructed.

However, when starting from a local free theory, one restricts the deformations
$\overscript{S}{(1)},\overscript{S}{(2)},...$ to be local functionals as well.
The above results do not take locality into account. If one imposes locality
of the
interacting action,
the analysis gets much more involved because the antibracket of two local BRST
cocycles
need not be in general the BRST variation of a local
functional\cite{Barnich,Barnich1}.
The obstructions that can arise have been illustrated in\cite{Barnich} in
the case of the
abelian Chern-Simons models, where the local BRST cohomology can be completely
analyzed
because the equations of motion are trivial and there is no local physical
degree of freedom.
In this note, we want to illustrate the usefulness of the cohomological
techniques
in the case of the deformation of free abelian vector fields.
We prove the uniqueness of the Yang-Mills coupling, recovering thereby in a
different fashion
results of\cite{Berends,Wald,Anco}.
The same techniques can be applied, in principle, to any other free gauge
theory.

\section{The Free Model}
\noindent
The action for several abelian vector fields in Minkowski space is given by
\begin{equation}
{\overscript{S}{(0)}}_0[A^a_\mu]=-{1\over 4}\int d^4x\
k_{ab}F^{\mu\nu a}F^b_{\mu\nu},\ \
F^a_{\mu\nu}=\partial_\mu A^a_\nu - \partial_\nu A^a_\mu
\end{equation}
where $k_{ab}$ is a non degenerate, symmetric and constant matrix.
This matrix must be positive definite in order for the physical Hamiltonian
to be bounded from below
and so we take $k_{ab}=\delta_{ab}$.
The equations of motion are $\partial^\nu F^a_{\nu\mu}=0$.
The minimal solution to the classical master equation is
\begin{equation}
\overscript{S}{(0)}={\overscript{S}{(0)}}_0+\int d^3x\
A^{\mu*}_a \partial_\mu C^a,
\end{equation}
leading to the BRST symmetry
\begin{equation}
\overscript{s}{(0)}=\partial_\nu F^{\nu\mu b}k_{ba}\overscript{
{\partial\over\partial A^{\mu*}_a}}{\rightarrow}
-\partial_\mu A^{\mu*}_a
\overscript{{\partial\over\partial C^*_{a}}}{\rightarrow}
+\partial_\mu C^a
\overscript{{\partial\over\partial A^{a}_\mu}}{\rightarrow}.
\end{equation}
One has ${\overscript{s}{(0)}}^2=0$, $[\overscript{s}{(0)},\partial_\mu]=0$ and
$\overscript{s}{(0)} d + d
\overscript{s}{(0)}=0$, where $d$ is the exterior spacetime derivative.

\section{Descent Equations and Construction of the Interaction Vertex}
\noindent
In this section, we determine all the local interaction vertices consistent to
first order in the coupling
constant $g$, i.e., we analyze eq.(\ref{deformation2}) with the further
requirement
that $\overscript{S}{(1)}$ should be a local functional.
Consistency to order $g^2$ (and higher)
is analyzed next.

The condition $\overscript{S}{(1)}=\int_{M^4}\overscript{\cal{L}}{(1)}\
\in H^0(\overscript{s}{(0)})$
implies for the integrand $\overscript{\cal{L}}{(1)}$
\begin{equation}
\overscript{s}{(0)}\overscript{\cal{L}}{(1)}+da_{[3]}=0\label{top}.
\end{equation}
Here, $\overscript{\cal{L}}{(1)}$
(respectively $a_{[3]}$) is a $4$-form (respectively 3-form) valued
polynomial in
the fields, the antifields and a finite number of their derivatives.
Furthermore, trivial interactions are eliminated by making the identification
$\overscript{\cal{L}}{(1)}\sim\overscript{\cal{L}}{(1)}+
\overscript{s}{(0)}b_{[4]}+dc_{[3]}$,
i.e., $\overscript{\cal{L}}{(1)}\in H^0(\overscript{s}{(0)}|d)$.
Eq.(\ref{top}) gives rise to a set of descent equations because of the
algebraic Poincar\'e
lemma\cite{Henneaux} :
\begin{eqnarray}
\overscript{s}{(0)}\overscript{\cal{L}}{(1)}+da_{[3]}&=0\label{topdescent}\\
\overscript{s}{(0)}a_{[3]}+da_{[2]}&=0\\
&\vdots\nonumber\\
\overscript{s}{(0)}a_{[b]}&=0\label{bottomdescent}
\end{eqnarray}
with $0\leq b \leq 4$ and $a_{[4]}\equiv\overscript{\cal{L}}{(1)}$.
As in\cite{Dubois-Violette} the analysis of eq.(\ref{top}) contains two steps:
\begin{romanlist}
\item First, one determines all possible inequivalent last elements $a_{[b]}$
of the descent consistent with Poincar\'e covariance. To simplify
the analysis, we impose the additional condition that the coupling constant $g$
has non positive dimension of length.
\item Second, one determines which last element $a_{[b]}$ can be lifted all
the way up to the
top of the ladder (\ref{topdescent}-\ref{bottomdescent}) to yield a solution of
eq.(\ref{top}). [As it is known, not all solutions of (\ref{bottomdescent})
come from a
descent, there may be obstructions].
\end{romanlist}
Both steps turn out to be immediate.

\noindent {\it Step 1:}

The last element $a_{[b]}$ of the descent has to be a d non-trivial
element of $H^g(\overscript{s}{(0)})$, with $g\geq 0$.
Taking into account the triviality of the Koszul-Tate differential $\delta$ at
positive antifield number and the results
of \cite{Brandt,Dubois-Violette1,Henneaux1}, one gets
\begin{equation}
a_{[b]}=a_{[b]}^\prime([F^a_{\mu\nu}],C^a)+s c_{[b]}
\end{equation}
where the notation $[\varphi^A]$ stands for $\varphi^A, \partial_\mu\varphi^A,
\partial_\mu\partial_\nu\varphi^A...$ . The BRST exact term corresponds to
trivial interactions
and can be dropped. Translation invariance imposes that $a_{[b]}^\prime$ should
not depend on
the coordinate $x^\mu$.
If $\overscript{\cal{L}}{(1)}$ is a cubic vertex, with ghost number zero and
form degree $4$, the
descent can have length at most three.
Furthermore, $a_{[D]}$ should have dimension (length) $-k$ with
$k\leq D$ and be of ghost number $4-D$. These requirements, together with the
condition
of Poincar\'e covariance, are easily seen to restrict non trivial descents to
have length $2$,
with last element given by
\begin{equation}
a_{[2]} = - {1\over 48} f_{abc} F^a C^b C^c\ \ {\rm or}\ \
a^\prime_{[2]} = - {1\over 48} f_{abc}\ ^*F^a C^b C^c.
\end{equation}
At this stage, $f_{abc}$ is required only to be antisymmetric in its last two
indices.

\noindent {\it Step 2:}

One can easily lift $a_{[2]}$ and $a^\prime_{[2]}$ once. This yields
\begin{equation}
a_{[3]} = - {1\over 24} f_{abc} F^a A^b C^c,\ \
a_{[3]}^\prime = - {1\over 24} f_{abc} \ ^*F^a A^b C^c +
{f^a}_{bc}\ ^* A^*_d C^b C^c.
\end{equation}
To lift $a_{[2]}$ or $a^\prime_{[2]}$ once more requires $f_{abc}$ to be
completely antisymmetric.
Otherwise $\overscript{\cal{L}}{(1)}$ does not exist. But if $f_{abc}$ is
completely antisymmetric,
then $a_{[2]}$ is BRST exact modulo $d$,
\begin{equation}
a_{[2]}=\overscript{s}{(0)} ( - {1\over 48} f_{abc}
A^a A^b C^c) + d (- {1\over 48} f_{abc}
A^a C^b C^c),
\end{equation}
and can be discarded.
Only $a^\prime_{[2]}$ defines non-trivial interactions, with
$\overscript{\cal{L}}{(1)}$
explicitly given by
\begin{eqnarray}
\overscript{\cal{L}}{(1)} =  f_{abc} ({1\over 2} F^{\mu \nu a}
{A_\mu}^b {A_\nu}^c
- A^{*\nu}_d k^{da}
{A_\nu}^b C^c-{1\over 2} C^*_d k^{da} C^b C^c ) d^4x.
\end{eqnarray}
The antifield independent part\footnote{As it is well known,
all the information about a BRST cocycle $A$ is contained in its antifield
independent part
$A_0$, of which $A$ is merely a ``BRST invariant extension". This is a standard
result of
Homological Perturbation Theory (see e.g.\cite{Henneaux}).}$\ $ of
$\overscript{\cal{L}}{(1)}$ is the standard Yang-Mills
cubic vertex except that, so far, $f_{abc}$ is not yet constrained to
fulfill the
Jacobi identity. The term linear in $ A^{*\mu}_a$ determines the modified
gauge transformations to
order $g$. Finally, the term linear in $ C^*_a$ shows that the modified
gauge transformations close
to order $g$ with structure constants equal to $f^a_{bc}$.

\section{Jacobi Identity}
\noindent
The existence of $\overscript{\cal{L}}{(1)}$ is equivalent to the consistency
of the interaction up to
order $g$. The interaction is then consistent also to order $g^2$ if and only
if
$\overscript{\cal{L}}{(2)}$ exists, i.e., if and only if one can satisfy
eq.(\ref{deformation3})
above with $\overscript{S}{(2)}=\int \overscript{\cal{L}}{(2)}$ a local
functional. This is where the
Jacobi identity comes in. Indeed,
in order to satisfy eq. (\ref{deformation3}) in the space of local
functionals, the
integrand of $(\overscript{S}{(1)},\overscript{S}{(1)})$, which is
\begin{eqnarray}
2( - {A^{*\mu}}_a {f^a}_{ec} C^c + F^{\mu \nu a} A_\nu^c f_{aec} +
\partial_\nu (A^{\mu a}
A_\nu^c) f_{eac}) ( -A_\mu^b C^d {f^e}_{bd})\nonumber\\
 + ({f^a}_{ce} C^*_a C^c + A^{*\mu}_a
{A^c}_\mu f_{ace}) {f^e}_{bd} C^b C^d
\end{eqnarray}
(up to a total divergence) must be equal to
$ - \overscript{s}{(0)} \overscript{\cal{L} }{(2)}  + \partial_\mu j^\mu$.

\noindent The term ${1\over 2} {f^a}_{ce} {f^e}_{bd} C^*_a C^c C^b C^d $
cannot be of that form, and hence must vanish.
This is the case if and only if the structure constants satisfy
the Jacobi identity. The terms of antifield number $1$ then also vanish
and
$\overscript{\cal{L}}{(2)}$ can be taken to be $-{1 \over 4}
A^{\mu a} A^{\nu c} {f^e}_{ac}
{A_\mu}^b  {A_\nu}^d f_{ebd}$, the well-known Yang-Mills quartic coupling.
The higher order equations are then satisfied with
$\overscript{\cal{L}}{(3)}=
\overscript{\cal{L}}{(4)}=...=0$. The
construction yields therefore the standard non-abelian
Yang-Mills models\footnote{It is
easy to see that are no quartic or higher order vertices
satisfying the above requirements.}$\ $ as unique renormalisable,
Poincar\'e covariant
deformations of several abelian vector
fields.

\section{Conclusions}
\noindent
In this paper, we have demonstrated the uniqueness of the Yang-Mills vertex.
We believe that
our approach is more interesting than the result itself, which has been
indeed already derived
in the literature\cite{Berends,Wald,Anco}.
We have shown how the existence of consistent couplings can be reformulated in
terms of
various cohomologies, namely, the cohomology of the free BRST differential
$\overscript{s}{(0)}$,
the cohomology of $d$, and the cohomology of  $\overscript{s}{(0)}$ modulo
$d$, for
which various calculational tools have been developed in the context of the
algebraic study of
anomalies.

\vfill
\eject

\nonumsection{Acknowledgements}
\noindent
This work has been supported in part by research funds from the FNRS and a
research contract with
the Commission of the European Communities.

\nonumsection{References}

\end{document}